\documentclass[aps,prl,nofootinbib,twocolumn,superscriptaddress,floatfix,notitlepage]{revtex4-1}
\pdfoutput=1

\usepackage{graphicx}
\usepackage{amsmath,amsfonts,amssymb}
\usepackage{color}
\usepackage[breaklinks,colorlinks,urlcolor=blue,citecolor=blue,linkcolor=magenta]{hyperref}
\usepackage{verbatim}
\usepackage{enumitem}
\usepackage{hyperref}
\usepackage{natbib}
\usepackage{xspace}

\newcommand{\be}{\begin{equation}} 
\newcommand{\ee}{\end{equation}}
\newcommand{\bea}{\begin{equation}\begin{aligned}} 
\newcommand{\eea}{\end{aligned}\end{equation}}

\def\lsim{\mathrel{\raise.3ex\hbox{$<$\kern-.75em\lower1ex\hbox{$\sim$}}}}
\def\gsim{\mathrel{\raise.3ex\hbox{$>$\kern-.75em\lower1ex\hbox{$\sim$}}}}

\begin{document}

\title{Interference Effects in $\mathbf{gg \to H \to Z \gamma}$ Beyond Leading Order}

\author{Federico~Buccioni}
\affiliation{\mbox{Physics Department, Technical University Munich, James-Franck-Strasse 1, 85748 Garching, Germany}} 
\author{Federica~Devoto}
\affiliation{Rudolf Peierls Centre for Theoretical Physics, Oxford University, Parks Road, Oxford OX1 3PU, UK}
\author{Abdelhak~Djouadi}
\affiliation{Departamento de F\'isica Te\'orica y del Cosmos, Universidad de Granada,
E-18071 Granada, Spain}
\author{John~Ellis}
\affiliation{Physics Department, King’s College London, Strand, London, WC2R 2LS, United Kingdom}
\affiliation{Theoretical Physics Department, CERN, CH 1211 Geneva, Switzerland}
\author{J\'er\'emie~Quevillon}
\affiliation{LAPTh, CNRS, USMB, 9 Chemin de Bellevue, 74940 Annecy, France}
\author{Lorenzo~Tancredi$^1$}

\begin{abstract}
The ATLAS and CMS collaborations at the LHC have recently announced evidence for the rare Higgs boson decay into a $Z$ boson and a photon. 
We analyze the interference between the process $gg\! \to \! H \! \to \! Z \gamma$ induced by loops of heavy particles, 
which is by far the dominant contribution to the signal, and the continuum $gg \to Z \gamma$ QCD background process mediated by light quark loops. 
This interference modifies the event yield, the resonance line-shape and the apparent mass of the Higgs boson. We calculate the radiative corrections 
to this interference beyond the leading-order approximation in perturbative QCD and find that, while differing numerically from the corresponding effects on the 
more studied $gg \! \to \! \gamma \gamma$ signal, they are generally rather small. As such, they do not impact significantly the interpretation of the 
present measurements of the $H \to Z \gamma$ decay mode.\\~~\\
KCL-PH-TH/2023-70, CERN-TH-2023-228, TUM-HEP-1488/23, OUTP-23-18P
\end{abstract}

\maketitle

\noindent\textbf{Introduction}\smallskip

{The study of the fundamental properties of the Higgs boson discovered in 2012 \cite{Aad:2012tfa,Chatrchyan:2012xdj} is high on the agenda of the LHC experimental collaborations. In particular, precise measurements of the Higgs boson production and decay rates in all accessible channels are of the utmost importance as they allow for the determination of the Higgs couplings to the known particles and probe possible effects of physics beyond the Standard Model (SM). A new frontier in Higgs physics has recently been opened by the} evidence for the rare decay mode into a $Z$ boson and a photon, $H\! \to\! Z \gamma$, that has been presented in a joint publication by the ATLAS and CMS collaborations~\cite{CMS:2023mku}, with a signal yield that is $\mu = 2.2 \pm 0.7$ times the rate expected in the SM. {The interpretation of this signal, and its potential excess, is of great topical interest.

The $H\! \to\! Z \gamma$ decay mode~\cite{Bergstrom:1985hp} (see also Ref.~\cite{Cahn:1978nz} for the reverse decay $Z\! \to\! H \gamma$) is of particular interest for seve\-ral reasons. It is unique among the Higgs decays observed to date in that the final state is neither a pair of identical particles (such as $\gamma \gamma$) nor a particle-antiparticle pair (such as $\bar b b$). Moreover, as a loop-induced process, the $H \to Z \gamma$ decay can yield a non-trivial check of the SM at the quantum level, providing constraints on the structure of the Higgs boson and its possible couplings to heavy particles \cite{Gunion:1989we,Djouadi:2005gi}.  This process is thus complementary to those from Higgs decays into two photons $H\! \to\! \gamma \gamma$ \cite{Ellis:1975ap} and the dominant gluon-fusion production mechanism $gg\! \to\! H$ \cite{Georgi:1977gs} which also proceed through loops of heavy particles: see, e.g., Refs.~\cite{Djouadi:1996yq,Chiang:2012qz,Chen:2013vi,Azatov:2013ura,Cao:2018cms}. For this reason, it is intriguing that the 3.4$\sigma$ signal reported by ATLAS and CMS has a strength that is somewhat higher than the SM value, leaving some space for possible physics beyond the SM, though such an apparent excess is not unexpected when a new process is first discovered.

In order to evaluate possible interpretations of the $pp\!  \to\!  H \!  \to\!  Z \gamma$ signal yield, it is important to have at hand the most accurate available calculation in the SM. {  
As in the more studied $\gamma\gamma$ channel \cite{Dicus:1987fk,Dixon:2003yb,Martin:2013ula,Dixon:2013haa,deFlorian:2013psa,Djouadi:2016ack,Campbell:2017rke,Bargiela:2022dla}, the $\gamma Z$ final state originates  not only from the signal process $gg \to H \to Z\gamma$ but also from the QCD background process $gg \to Z\gamma$ which proceeds through box diagrams involving SM light quarks. The interference of the two  processes could  modify not only the signal rate but also the resonance line-shape and the apparent mass of the Higgs boson.}

In this paper, we present the results of a calculation of the dominant radiative corrections at next-to-leading order (NLO) in perturbative QCD of the signal strength in the SM, including the interference with the QCD background $gg\to Z\gamma$. For the latter, we take into account the virtual corrections generated by light-quark loops in two-loop diagrams and the real corrections with soft-gluon emission in one-loop box and pentagon diagrams. We find that these effects differ numerically from the corresponding NLO effects in the well-known and more studied $gg\! \to \! \gamma \gamma$ process. The NLO corrections are not large. Consequently, they do not impact significantly the apparent tension between the SM prediction and the recent ATLAS and CMS measurements. We therefore await with interest the evolution of this measurement with the accumulation of LHC luminosity.
\\

\noindent\textbf{The $\mathbf{gg \to Z\gamma}$ process}\smallskip

At leading-order (LO), the SM diagram for the dominant Higgs
production mechanism at the LHC, $gg \to H$ \cite{Georgi:1977gs}, followed by the $H \to Z \gamma $ decay \cite{Bergstrom:1985hp}, is shown in the left panel of Fig.~\ref{fig:feynmanpp}. As far as the dominant QCD interactions are concerned, the higher-order corrections to $gg\to H$ production have been given at NLO in Refs.~\cite{Djouadi:1991tka,Dawson:1990zj,Spira:1995rr} and have been calculated up to N$^3$LO in the very large top mass limit, $2m_t \gg M_H$ \cite{Anastasiou:2015vya,Chen:2021isd}. The corrections are large, increasing the LO cross section by more than a factor of two. In the case of the $H\! \to \! Z\gamma$ decay, the NLO QCD corrections to the relevant top-quark loop have been found to be very small \cite{Spira:1991tj,Gehrmann:2015dua,Bonciani:2015eua}. We note, moreover, that in the SM the $W$-boson loop contribution dominates by far over the top-loop contribution, while those of the $b$-quark and other fermions are negligible. 

\begin{figure}[!h]
\vspace*{-0.6cm}
\centerline{ \includegraphics[scale=.6]{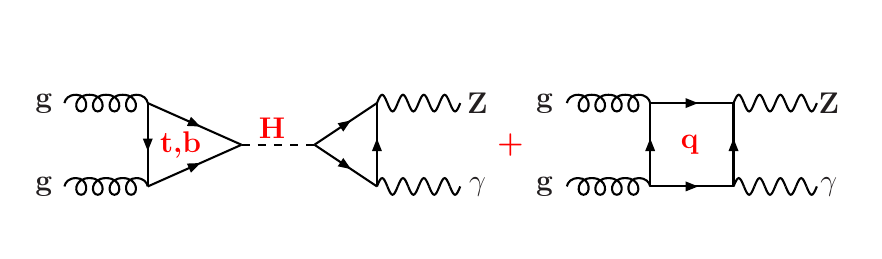} }
\vspace*{-.7cm}
\caption{Feynman diagrams at leading order for the signal (left) and the background (right) in the process $gg\to Z \gamma$.} 
\label{fig:feynmanpp}
\end{figure}
The $Z\gamma$ final state is also produced in pure QCD in gluon fusion, $gg\to Z\gamma$, through the box diagram  shown in the right panel of Fig.~\ref{fig:feynmanpp}. The cross section for this background process, which is mediated by light quark loops as heavy quarks decouple at large masses, has been given at LO in Refs.~\cite{Ametller:1985di,vanderBij:1988fb,Adamson:2002rm}. (We note that there is also a very large contribution to the background from the $q\bar q \! \to \! \gamma Z$ tree-level process, but we ignore it as it does not contribute to the interference which is our main concern here.) At NLO, the QCD corrections are analogous to those derived in the $gg\!\to \! \gamma\gamma$ case \cite{Bern:2001df,Catani:2011qz,Campbell:2016yrh} and the relevant two-loop helicity amplitudes have been given in Ref.~\cite{Gehrmann:2013vga}. Note that, at NLO, we assume  massless quarks in the background process, whereas the infinite top quark mass limit is assumed in the Higgs signal process.

As the Higgs signal and continuum background processes have the same initial and final states, they will interfere. The total amplitude for the $gg\to Z\gamma$  process including both contributions may be then written as
\begin{equation}
{\cal A} = -  \frac{{\cal A}_{gg H} {\cal A}_{HZ\gamma}}{{ M}_{Z \gamma}^2 - M_H^2 + i M_H \Gamma_H} + {\cal A}_{gg Z \gamma } \, ,
\label{eq:ggpp}
\end{equation}
where ${M}_{Z \gamma}$ is the invariant mass of the $Z \gamma$ system and, in the first term, the denominator represents the propagator of the Higgs boson with mass $M_H$ and total width $\Gamma_H$, which we take to be $M_H\!=\!125$ GeV and $\Gamma_H\!=\!4.07$ MeV~\cite{ParticleDataGroup:2022pth}. The differential cross section for the  full $gg\to \gamma Z$ process including the Higgs signal, the continuum background and their interference is given schematically by
\begin{equation}
\frac{{\rm d}\sigma (gg \! \to  \! \gamma Z) }{  
\, {\rm d} { M}_{Z \gamma} } %
\propto
 \frac {N_S \!+\! N_I^{\rm Re} \!+\! N_I^{\rm Im} } { ( { M}_{Z \gamma}^2 \!-\! M_H^2)^2 \!+\! M_H^2 \Gamma^2_H} \!+\! N_B  \, ,
\label{eq:sigmall}
\end{equation}
where the various components read
\begin{eqnarray}
N_S &=& |{\cal A}_{gg H } \, {\cal A}_{HZ\gamma}|^2 \, ,   \ \ 
N_B= |{\cal A}_{gg Z \gamma}|^2 \, ,  \\
N_I^{\rm Re} &=&   -2 {\rm Re} [ {\cal A}_{gg H} \, {\cal A}_{HZ\gamma} \,  {\cal A}_{ggZ\gamma }^* ] \times ({ M}_{Z \gamma}^2 - M_H^2) \, ,  \\
N_I^{\rm Im} &=&  -  2 {\rm Im} [ {\cal A}_{gg H} \, {\cal A}_{HZ \gamma} \, {\cal A}_{ggZ\gamma}^* ] \times M_H \Gamma_H \, . 
\end{eqnarray}
Here, we refrain from giving explicit expressions for the various amplitudes and simply describe the relevant contributions.
For the interference, its first component $N_I^{\rm Re} \propto {M}_{Z \gamma}^2 \!-\! M_H^2$ does not  contribute to the cross section when integrating over the invariant mass, as the gluon luminosity varies slowly over the total width $\Gamma_H$. However, it distorts the resonance shape and shifts the position of the peak, changing the apparent mass of the resonance. On the other hand, the second interference term $N_I^{\rm Im}$ contributes to the cross section, though its contribution is suppressed by the small Higgs width $\Gamma_H$.  

The interference $N_I^{\rm Im}$ requires  an absorptive part in the amplitudes. For the light quarks with $m_q \ll M_{Z,H}$, the contribution of mass corrections to the imaginary part  is suppressed by powers of $4m_q^2/ M^2_{Z,H}$ relative to the value in the massless limit. The amplitudes for the induced $gg H$ and $HZ\gamma$ couplings involve imaginary components when the particles circulating in the loops have masses below the kinematical threshold, i.e., $ M_H^2 \geq 4m_X^2$, which is the case for bottom quarks. In the case of the $HZ\gamma$ amplitude, the $b$--quark contribution is completely negligible. In turn, the $b$-quark contribution to the $Hgg$ amplitude is sizeable and its interference with the dominant top contribution represents about 10\% of the total amplitude (at LO) and it has a non-negligible imaginary component \cite{Gunion:1989we,Djouadi:2005gi}.

{Since the focus of this paper is to assess whether the interference can resolve the tension between the SM prediction and the $H \to Z\gamma$ signal rate recently measured at the LHC, we consider those contributions that could be responsible for an enhancement of the interference effects which go beyond naive expectations from perturbative QCD. Such effects are usually given by strong phases, e.g., absorptive contributions arising from light on-shell particles in the loops. It is clear that such phases, if present, would come from virtual corrections, and are therefore correctly taken into account by the soft-virtual approximation.}}\\

\noindent\textbf{The $gg \to Z\gamma$ interference at NLO}\smallskip 

\begin{figure}[!h]
\vspace*{-0.6cm}
\centerline{
\includegraphics[scale=0.55]{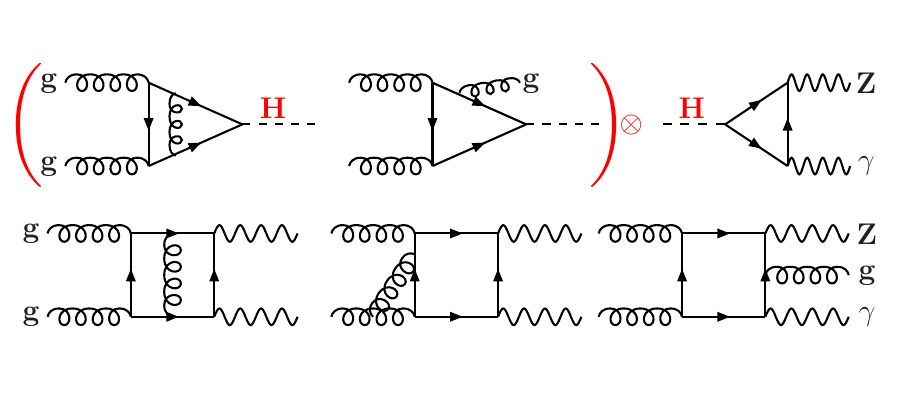} }
\vspace*{-0.8cm}
\caption{Generic Feynman diagrams in NLO QCD for the signal process $gg\to H$ with the decay $H\to Z\gamma$ (upper row) and for the background process $gg\to Z \gamma$ (lower row).} 
\label{fig:2loop}
\vspace*{-2mm}
\end{figure}

At next-to-leading-order in perturbative QCD, some generic Feynman diagrams that contribute to the $gg\to Z \gamma$ process are given in Fig.~\ref{fig:2loop}. The upper row shows some NLO diagrams for the signal process $gg\to H$ with the subsequent decay $H\to Z\gamma$, and the lower row shows some of the NLO ones that contribute to the SM background. The latter involve two-loop box diagrams with virtual gluon exchange and one-loop diagrams (including pentagons) accompanied by real gluon emission.

We report in this paper the results of a simplified SM calculation at NLO in perturbative QCD that captures the most important features of the interference effects in this $gg \to Z\gamma$ search channel.
We treat the interference term in the so-called \textit{soft-virtual} approximation, i.e.,
we retain the complete contribution from virtual corrections, but neglect the impact of hard QCD radiation and account only for soft real emissions~\cite{Kramer:1996iq,Catani:2001ic,Catani:2003zt,deFlorian:2012za,Beneke:2019mua}. A brief description of the
soft-virtual approximation is in order. Here, we outline only those features that are relevant for the discussion of our results. For a detailed derivation we refer the reader to~\cite{deFlorian:2012za} and to~\cite{Bargiela:2022dla} where the application to the $\gamma\gamma$ interference case is discussed.
We define $z=Q^2/\hat{s}$, where $Q$ is the invariant mass of the produced final state, 
in our case the $Z\gamma$ system, and $\sqrt{\hat{s}}$ is the partonic center-of-mass energy.
The soft-virtual approximation amounts to an expansion of the partonic cross section around the $z\to1$ limit. In this limit the partonic cross section factorises as
\begin{align}
  &\mathrm{d}\hat\sigma\left(z,\lbrace\hat{x}_i\rbrace,\alpha_s,\frac{Q^2}{\mu^2_R},\frac{Q^2}{\mu^2_F}\right) \simeq \notag \\
  &\mathrm{d}\hat\sigma_{\rm{LO}}\left(\lbrace\hat{x}_i\rbrace,\alpha_s\right) \, z \,
  G\left(z,\alpha_s,\frac{Q^2}{\mu^2_R},\frac{Q^2}{\mu^2_F}\right),
  \label{eq:softvirtapprox}
\end{align}
where $d\hat\sigma_{\rm{LO}}$ is the LO contribution and $\lbrace\hat{x}_i\rbrace$ a generic set of variables describing fully the final-state dynamics. The function $G$ is then expanded perturbatively in $\alpha_s$,
\begin{equation}
G\left(z,\alpha_s\right) = \delta(1-z) + \sum_{n=1} \left(\frac{\alpha_s}{2\pi}\right)^n 
G^{(n)}\left(z\right),
\end{equation}
where we have suppressed the dependence on the scales $\mu_{R,F}$ which is implicit in the coefficients $G^{(n)}(z)$. These are the dominant terms in the $z\to 1$ limit, which are given by $\delta(1-z)$ and standard plus distributions 
\begin{equation}
\mathcal{D}_n(z) = \left[ \frac{\ln^n(1-z)}{1-z}\right]_{+}\, .
\end{equation}
Various refinements to the soft-virtual approximation have been proposed in the literature, 
especially in the case of resonant final states such as 
$Z/W$ and Higgs production in the infinite top-mass limit~\cite{Catani:2001ic,Moch:2005ky,Ball:2013bra,deFlorian:2014vta}. However, the picture is not equally well established
for processes induced by a loop of light quarks such
as the continuum background.

Here, in order to provide a more reliable estimation of the uncertainties in our calculation, we also consider an alternative approach to the \textit{naive} soft-virtual approximation of Eq.~\eqref{eq:softvirtapprox}.
We follow the proposal in~\cite{Ball:2013bra}, where 
subleading terms in the soft expansion are partially captured by replacing
\begin{equation}
\label{eq:softvirtv2}
    \mathcal{D}_n(z) \to \mathcal{D}_{n}(z) + (2-3z+2z^2)\frac{\ln^n\frac{1-z}{\sqrt{z}}}{1-z} - 
    \frac{\ln^n(1-z)}{1-z}\,.
\end{equation}
This method has also been adopted for treating the production of a pair of $W$ bosons at high energy~\cite{Bonvini:2013jha}.
Although there is no compelling reason why Eq.~\eqref{eq:softvirtv2} should provide more reliable results than a naive soft-virtual expansion for the interference, 
we use the difference between these two predictions as a way to estimate
our theory uncertainty at NLO.
We stress however that the SM signal cross section is, instead, treated exactly through NLO in QCD, i.e., retaining the full dependence on the real radiation. 

Our NLO QCD prediction relies on the two-loop helicity amplitudes for the background process presented in Ref.~\cite{Gehrmann:2013vga}.
The full helicity information allows us to incorporate the spin-correlated decay of the $Z$ boson. In this paper we consider 
in particular its electron-positron
decay channel, $Z\to e^- e^+$. In principle, the complete $pp\to e^- e^+ \gamma$ scattering process also includes a contribution where an off-shell
photon decays to leptons, $pp\to \gamma^{*}(e^- e^+)\gamma$, and one where the $Z$ boson decays to an $e^-e^+$ pair and the final-state photon is emitted off the leptons.
However, both effects are negligible for the particular analysis we present in this paper. The impact of the former is significantly reduced by the selection cuts we impose on the final-state leptons. 
The latter is expected to be small based on the fact
that we focus on the Higgs-boson resonance region, which forces the invariant mass of the $e^-e^+\gamma$ system to be away from the $Z$ resonance peak.

In choosing the selection criteria for the leptons and the photon, we were inspired by an  analysis of the $H\! \to \! Z\gamma$ channel carried out by the ATLAS collaboration~\cite{ATLAS:2020qcv},
and adopted analogous cuts used for the $Z\to e^+e^-$ decay mode. These cuts are rather inclusive and yield results that are qualitatively similar to the inclusive case. The difference is restricted to an overall normalisation and simply results in a small reduction of the cross section.
Specifically, we require the $e^-e^+$ pair to have an invariant mass $m_{\ell\ell}$ in the range $50\;\mathrm{GeV}<m_{\ell\ell}<101\;\mathrm{GeV}$,
that all tagged final-state particles satisfy $p_{T,i} > 10 \; \mathrm{GeV}$, with $i \in \lbrace e^-,e^+,\gamma\rbrace$, and finally that the rapidities are constrained by $|y_{e^{\pm}}|<2.47$ and $|y_{\gamma}|<2.37$.

The setup of our calculation is as follows. We adopt the $G_\mu$ scheme for the electroweak parameters and choose $M_Z = 91.1876\;\mathrm{GeV}$,
$M_W = 80.398\;\mathrm{GeV}$, $\Gamma_Z = 2.4952\;\mathrm{GeV}$ and $G_F = 1.16639\cdot 10^{-5}\;\mathrm{GeV}^2$, which results in $\alpha = 1/132.277$.
For the Higgs boson mass we choose $M_H = 125\;\mathrm{GeV}$.
Our predictions are derived for a $pp$ collider at a centre-of-mass energy of $\sqrt s =13.6$ TeV, i.e. the current LHC operational mode.
We use the \texttt{NNPDF31\_nlo\_as\_0118} set~\cite{NNPDF:2017mvq} of parton distribution functions (PDFs) with the value $\alpha_s(M_Z)=0.118$,
and we make use of the $\texttt{LHAPDF}$~\cite{Buckley:2014ana} and $\textsc{Hoppet}$~\cite{Salam:2008qg} programs for manipulation of the PDFs. 
For the QCD factorisation and renormalisation scales, we have chosen a common reference value $\mu_F=\mu_R=\mu_0 = \frac12 M_H$. Theory uncertainties for the signal process and for the LO interference are estimated by a simultaneous rescaling of the nominal value by factors $2$ and $1/2$. As for the interference at NLO, as described above,
the spread is defined by the difference in the standard soft-virtual approximation and its modification described in Eq.~\eqref{eq:softvirtv2}.

\begin{figure}[t]
    \centering    \includegraphics[angle=270,width=\columnwidth]{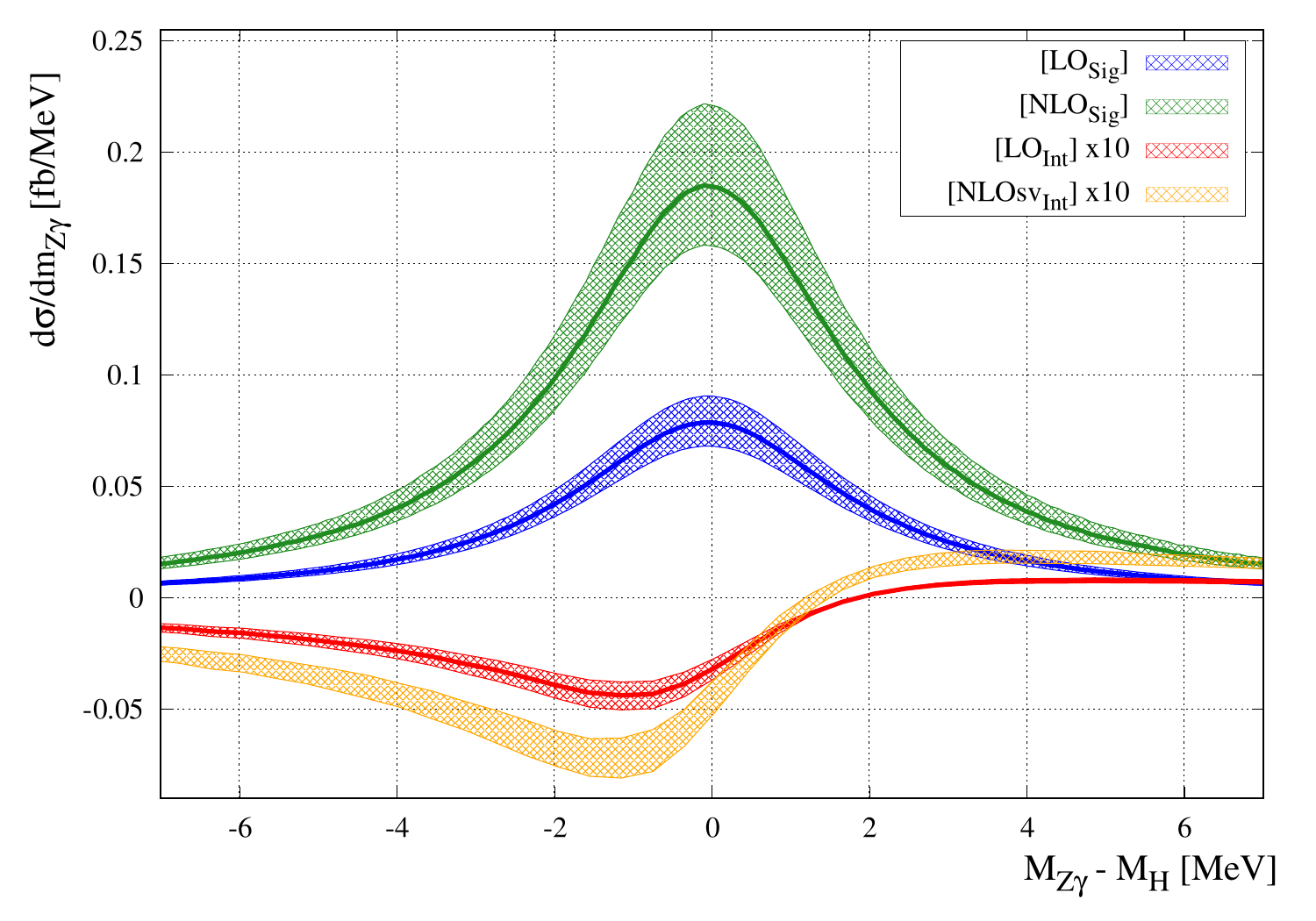}
\vspace*{-2mm}
    \caption{ The $gg\! \to\! H \to\! Z \gamma$ signal cross section at LO (in blue) and at NLO in QCD (in green), and the interference with the $gg\!\to\! \gamma Z$ QCD background at LO (in red) and at NLO but in the soft-virtual approximation 
    (in orange). The bands represent the scale variation, except for $\mathrm{NLO_{SV}}$ where
    the band shows the spread between two different soft-virtual approximations, see the text for details.  The results are for the LHC with $\sqrt s\!=\! 13.6$~TeV, and the interference terms are magnified by a factor of 10 for visualisation purposes.}
        \label{fig:column}
\vspace*{-2mm}
\end{figure}

Fig.~\ref{fig:column} displays our main findings.  It shows the signal and interference (magnified by a factor of 10) line-shapes at LO and NLO as functions of the difference between the $Z \gamma$ invariant mass and the Higgs boson mass. 
The central values of the cross sections are obtained for the reference scale choice, while the bands stem from scale variations.
The LO cross section for the signal, $gg \to H \to Z(e^-e^+) \gamma$, is shown as a blue band, and the NLO cross-section is shown as a green band. The ratio between the two is the well-known $K$ factor of about $\simeq 2$ \cite{Djouadi:1991tka,Dawson:1990zj,Spira:1995rr}.  
The red band shows the result of the calculation of the signal-background interference at LO and the orange band is the result of
our NLO interference calculation in the soft-virtual approximation, i.e., including virtual contributions and the leading real contributions from soft-gluon emission. 

We estimate the uncertainty band of the interference contribution from the spread between the two different approaches to the soft-virtual approximations discussed above. We point out that the standard scale-variation uncertainty on  the interference term (obtained by dividing and multiplying the central scale by a factor of two) is accidentally small and is contained within our more conservative estimate.

We note that, already at LO, the interference tends to reduce the total cross section and shift the effective Higgs mass to higher values. However, both these effects are numerically very small (recall the factor of 10 of magnification in Fig.~\ref{fig:column}).
The effects at NLO are qualitatively similar to those at the LO, and are larger numerically by a small factor that depends on the value of the invariant mass of the $Z\gamma$ system relative to the Higgs pole mass, as can be seen in Fig.~\ref{fig:column}. This differs from the large $K$ factor for the non-interference term.

We observe that, if we restrict the invariant mass window of the $Z\gamma$ system to a very narrow window, $124-126\;\mathrm{GeV}$, the cross sections of 
the signal and interference terms in the fiducial volume outlined above are
\begin{equation}
    \label{eq:xsectsresults}
    \sigma^{\mathrm{NLO}}_{\mathrm{Sig}} \!=\! 1.207^{+20\%}_{-15\%} \; \mathrm{fb}, \    \sigma^{\mathrm{NLO_{SV}}}_{\mathrm{Int}} \!=\! -0.0344^{+12\%}_{-12\%} \; \mathrm{fb},
\end{equation}
where the label SV refers to the prediction in the soft-virtual approximation, and its
central value is the mean of the results extracted in the two approaches described above.
The uncertainty is assessed as discussed for Fig.~\ref{fig:column}.
Thus, we estimate that the interference has a destructive impact on the total rate of $\mathcal{O}(-3\%)$\footnote{This number may be subject to small variations upon including large missing higher-order QCD corrections in the signal process.}.

The NLO QCD corrections  to the interference in the $gg\to Z\gamma$ process are small and do not modify substantially the signal rate in the Higgs $gg \to H \to Z \gamma$ production channel. These effects also do not modify significantly the effective Higgs mass measured in the $Z \gamma$ final state, which should be indistinguishable from that measured in the $\gamma \gamma$ and $ZZ$ final states.

As a final remark, it is worth mentioning a few qualitative differences with respect to the more deeply investigated interference in the $\gamma\gamma$ decay channel. In the latter, the Higgs boson decays to a pair of massless spin-1 particles, and given the scalar nature of the Higgs, these photons must have identical helicities~\cite{Dicus:1987fk,Martin:2013ula}. In this configuration, the corresponding LO background amplitudes receive an imaginary contribution that is suppressed by the ratio $m^2_q/M_H^2$ where $m_q$ is the mass of a light quark running in the loop, see, e.g., Fig.~\ref{fig:feynmanpp}. Therefore, a noticeable destructive effect arises only at NLO~\cite{Dixon:2003yb}.
In the $Z\gamma$ mode instead, such a helicity selection does not occur and an absorptive part
is already manifest at LO. Furthermore, the real contribution $N_I^{\mathrm{Re}}$ has an opposite impact on the line shape than the one in $\gamma\gamma$, which induces a slight excess of events to right of the Higgs boson peak and not to the left as in $\gamma\gamma$.
\\

\noindent\textbf{Conclusions}\smallskip

We have considered the interference between the signal amplitude for Higgs production at the LHC and its subsequent decay into a photon and a $Z$ boson, $gg\to H\to Z\gamma$, and the amplitude for the pure QCD background process $gg\to Z\gamma$ beyond leading order in perturbative QCD.
The conclusions from our analysis are that interference effects beyond leading order are small given the current
experimental accuracy, as was already the case for the leading-order QCD interference effects. They do not modify significantly the apparent tension between the SM prediction for the $gg \to H \to Z \gamma$ signal strength and that recently measured by the ATLAS and CMS collaborations, namely $\mu = 2.2 \pm 0.7$. 

As such, they leave space for alternative explanations of the apparent tension: either the speculative possibility of physics beyond the SM (see, e.g.,~\cite{Boto:2023bpg}) or, more plausibly, a statistical fluctuation. We await with interest the accumulation of more LHC data to resolve this issue.

\begin{acknowledgments}
\vspace{5pt}\noindent\emph{\bf Acknowledgments}\smallskip 

We are grateful to Fabrizio Caola for valuable discussions and for comments on the manuscript.
AD is supported by the Junta de Andalucia through the Talentia Senior program and the grant PID2021-128396NB-I00. The work of JE was supported in part by STFC Grant ST/X000753/1. This work was supported in part by the Excellence Cluster ORIGINS funded by the Deutsche Forschungsgemeinschaft (DFG, German Research Foundation) under Germany’s Excellence Strategy – EXC-2094-390783311 
and by the European Research Council (ERC) under the European Union research and innovation programme grant agreements 949279 (ERC Starting Grant HighPHun) and 804394 (ERC Starting Grant \textsc{hipQCD}).
\end{acknowledgments}
\vspace{-0.6cm}
\bibliography{refs}

\end{document}